\documentclass[a4paper,12pt]{article}

\begin{document}

\author{C. Barrab\`es\thanks{E-mail : barrabes@celfi.phys.univ-tours.fr}\\     
\small Laboratoire de Math\'ematiques et Physique Th\'eorique\\
\small  CNRS/UPRES-A 6083, Universit\'e F. Rabelais, 37200 TOURS, France\\
P.A. Hogan\thanks{E-mail : phogan@ollamh.ucd.ie}\\
\small Mathematical Physics Department\\
\small  National University of Ireland Dublin, Belfield, Dublin 4, Ireland}

\title{Braking--Radiation: An Energy Source for a Relativistic Fireball}
\date{}
\maketitle

\begin{abstract}
If the Schwarzschild black--hole is moving rectilinearly with 
uniform 3--velocity and suddenly stops, 
according to a distant observer, then we demonstrate that 
this observer will see a spherical light--like shell or ``relativistic fireball" 
radiate outwards with energy equal to the 
original kinetic energy of the black--hole.

\end{abstract}
\thispagestyle{empty}
\newpage

\indent
Current models for the source of gamma ray bursts provide a 
strong motivation to explore, in the context of General Relativity, 
the construction of models of light--like spherical shells of 
matter (``relativistic fireball''). 
In the fireball model of gamma ray bursts a relativistically 
expanding shell is slowed down by the interstellar medium and 
its energy is converted to gamma rays \cite{PIR}.
 Currently proposed sources are binary neutron 
star mergers \cite{EIC}, failed supernovae \cite{WOO} and 
magnetic white dwarf collapse \cite{U}. In this paper we propose a new source 
for a relativistic fireball. A Schwarzschild black--hole 
moving rectilinearly with 
uniform 3--velocity according to a distant observer  
suddenly stops. We show that this results in  
a spherical light--like shell or relativistic fireball 
propagating outwards. The total energy of the fireball 
measured by the distant observer is the same as the 
original kinetic energy of the black--hole. In our model the 
mechanism for the production of the shell is not considered. 
However the effect of the shell in producing zero recoil velocity 
in the black--hole aswell as the origin of the total energy of 
the shell is manifest. 

We use the Schwarzschild line--element in the form
\begin{equation} \label{1}
ds^2=k^2\,r^2\left\{\frac{d\xi ^2}{1-\xi ^2}+
(1-\xi ^2)\,d\phi ^2\right\}-2\,du\,dr-
\left (1-\frac{2\,m}{r}\right )\,du^2\ ,
\end{equation}
with $k^{-1}=\gamma\,(1-v\,\xi )$ , $v$ is a real constant 
such that $0<v<1$ and $\gamma =(1-v^2)^{-1/2}$. We can remove 
the parameter $v$ from (\ref{1}) and recover the usual form 
of the Schwarzschild line--element by making the replacement 
\begin{equation} \label{2}
\xi\longrightarrow \frac{v+\xi}{1+v\,\xi}\ .
\end{equation}
The inverse of this transformation is a Lorentz boost in the 
direction $\xi =+1$ viewed by a 
distant observer \cite{BREM}. In this space--time $u={\rm constant}$ are null 
hypersurfaces generated by the geodesic integral curves of the 
(future--pointing) null vector field $\partial /\partial r$, 
with $r$ an affine parameter along them. These null hypersurfaces 
are future--directed null--cones in the sense that the generators 
are geodesic, shear--free and have expansion $r^{-1}$.

We now subdivide the space--time with line--element (\ref{1}) 
into two halves $M^-$ and $M^+$, each with boundary the 
future null--cone ${\cal N} (u=0)$. To the past $(u<0)$ of ${\cal N}$ 
the space--time $M^-$ has line--element (\ref{1}). To the 
future $(u>0)$ of ${\cal N}$ the space--time $M^+$ has line--element
\begin{equation} \label{3}
ds_+^2=r_+^2\left\{\frac{d\xi _+^2}{1-\xi _+^2}+
(1-\xi _+^2)\,d\phi _+^2\right\}-2\,du\,dr_+-
\left (1-\frac{2\,m_+}{r_+}\right )\,du^2\ .
\end{equation}
The space--times $M^-$ and $M^+$ are attached on ${\cal N}$ with 
the matching conditions
\begin{equation} \label{4}
\xi =\xi _+\ ,\phi =\phi _+\ ,r=r_+k^{-1}\ ,
\end{equation}
with
\begin{equation} \label{5}
k^{-1}=\gamma\,(1-v\,\xi _+)\ .
\end{equation}
These conditions ensure that the metric on ${\cal N}$ induced by its embedding 
in $M^-$ is the same as the metric on ${\cal N}$ induced by its 
embedding in $M^+$. A detailed motivation for this matching 
is given in \cite{BREM}. The physical picture is as follows: 
relative to the observer using the plus coordinates 
a Schwarzschild black--hole moving rectilinearly with uniform 
3--velocity $v$ has its 3--velocity suddenly reduced to zero 
and this is followed by the emergence of a spherical--
fronted light--like signal. For greater generality we have 
assumed that the rest--mass of the black--hole has changed 
from $m$ in $M^-$ to $m_+$ in $M^+$. We now consider the physical properties of the signal 
with history the future null--cone ${\cal N}$. To do this we use the theory of 
light--like signals in general relativity  developed
by  Barrab\`es--Israel (BI) \cite{BI}. The BI theory enables 
us to calculate, if it exists, the coefficient of $\delta(u)$ 
in the Einstein tensor of $M^-\cup M^+$. This coefficient, if 
non--zero, is simply related to the surface stress--energy tensor 
of a light--like shell with history ${\cal N}$. The theory also 
enables us to calculate the coefficient of $\delta (u)$ 
in the Weyl tensor of $M^-\cup M^+$ if it exists. This allows 
us to determine whether or not the light--like signal with history 
${\cal N}$ includes an impulsive gravitational wave \cite{BBH}. For 
the details of the BI technique the reader must consult 
\cite{BI} and further developments are to be found in \cite{BH}. 
We will merely guide the reader through the present application 
of the theory.
The local coordinate system in $M^-$ with line--element (\ref{1}) 
is denoted $\{x^\mu _-\}=\{\xi , \phi , r, u\}$ while the local 
coordinate system in $M^+$ with line--element (\ref{3}) is 
denoted $\{x^\mu _+\}=\{\xi _+, \phi _+, r_+, u\}$. The equation 
of ${\cal N}$ is $u=0$ and thus we take as normal 
to ${\cal N}$ the null vector field with components $n_\mu$ given via 
the 1--form $n_\mu dx^\mu _{\pm}=-du$. Since we wish to 
discover the physical properties of ${\cal N}$ observed by 
the observer using the plus coordinates, we take 
$\{\xi ^a\}=\{\xi _+, \phi _+, r_+\}$ with $a=1, 2, 3$ as 
intrinsic coordinates on ${\cal N}$. 
A set of three linearly independent tangent vector fields to ${\cal N}$ 
is $\left\{e_{(1)}=
\partial /\partial\xi _+, e_{(2)}=\partial /\partial\phi _+, 
e_{(3)}=\partial /\partial r_+\right\}$. The components of these 
vectors on the plus side of ${\cal N}$ are 
$e^\mu _{(a)}|_+=\delta ^\mu _a$. The components of these 
vectors on the minus side of ${\cal N}$ are 
\begin{equation} \label{6}
e^\mu _{(a)}|_-=\frac{\partial x^\mu _-}{\partial\xi ^a}\ ,
\end{equation}
with the relation between $\{x^\mu _-\}$ and $\{\xi ^a\}$ 
given by the matching conditions (\ref{4}). Hence 
we find that
\begin{eqnarray} \label{7}
e^\mu _{(1)}|_- &=& (1, 0, -r_+\gamma\,v, 0)\ ,\\
e^\mu _{(2)}|_- &=& (0, 1, 0, 0)\ ,\\
e^\mu _{(3)}|_- &=& (0, 0, \gamma\,(1-v\,\xi _+), 0)\ .
\end{eqnarray}
We need a transversal on ${\cal N}$ consisting of a vector field on 
${\cal N}$ which points out of ${\cal N}$. A convenient such (covariant) vector 
expressed in the coordinates $\{x^\mu _+\}$ is ${}^+N_\mu =
(0, 0, 1, \frac{1}{2}-\frac{m_+}{r_+})$. Thus since $n^\mu =\delta ^\mu _3$ 
we have ${}^+N_\mu n^\mu =+1$.We next construct the transversal 
on the minus side 
of ${\cal N}$ with covariant components ${}^-N_\mu$. To ensure that 
this is the same vector on the minus side of ${\cal N}$ as ${}^+N_\mu$ 
when viewed on the plus side we require
\begin{equation} \label{10}
{}^+N_ \mu\,e^\mu _{(a)}|_+={}^-N_ \mu\,e^\mu _{(a)}|_-\ ,
\qquad {}^+N_\mu {}^+N^\mu ={}^-N_\mu {}^-N^\mu\ .
\end{equation}
The latter scalar product is zero as we have chosen to 
use a null transversal. We find that
\begin{equation} \label{11}
{}^-N_\mu =\left (\frac{r_+v}{1-v\,\xi _+}, 0, \frac{1}
{\gamma\,(1-v\,\xi _+)}, D\right )\ ,
\end{equation}
with
\begin{equation} \label{111}
D=\frac{v^2(1-\xi _+^2)\,\gamma}{2(1-v\,\xi _+)}+
\frac{1}{2\gamma\,(1-v\,\xi _+)}-\frac{m}
{\gamma ^2(1-v\,\xi _+)^2r_+}\ .
\end{equation}
Next the transverse extrinsic curvature on the plus and minus 
sides of ${\cal N}$ is given by
\begin{equation} \label{12}
{}^{\pm}{\cal K}_{ab}=-{}^{\pm}N_\mu\left (\frac{\partial e^\mu 
_{(a)}|_{\pm}}{\partial\xi ^b}+{}^{\pm}\Gamma ^\mu _{\alpha\beta}\,
e^\alpha _{(a)}|_{\pm}e^\beta _{(b)}|_{\pm}\right )\ ,
\end{equation}
where ${}^{\pm}\Gamma ^\mu _{\alpha\beta}$ are the components 
of the Riemannian connection associated with the metric tensor of 
$M^+$ or $M^-$ evaluated on ${\cal N}$. The key quantity we need is 
the jump in the transverse extrinsic curvature across ${\cal N}$ given 
by
\begin{equation} \label{13}
\sigma _{ab}=2\,\left ({}^+{\cal K}_{ab}-{}^-{\cal K}_{ab}\right )
\ .
\end{equation}
This jump is independent of the choice of transversal on ${\cal N}$ 
\cite{BI}. We find that in the present application $\sigma _{ab}=0$ 
except for 
\begin{equation} \label{14}
\sigma _{11}=\frac{2}{1-\xi _+^2}\,\left (m\,k^3-m_+\right )\ ,
\qquad \sigma _{22}=2\,(1-\xi _+^2)\left (m\,k^3-m_+\right )\ ,
\end{equation}
with $k$ given by (\ref{5}). 
Now $\sigma _{ab}$ is extended to a 4--tensor field on ${\cal N}$ with 
components $\sigma _{\mu\nu}$ by padding--out with zeros (the 
only requirement on $\sigma _{\mu\nu}$ is $\sigma _{\mu\nu}\,
e^\mu _{(a)}|_{\pm}\,e^\nu _{(b)}|_{\pm}=\sigma _{ab}$). With our 
choice of future--pointing normal to ${\cal N}$ and past--pointing 
transversal, the surface stress--energy tensor components are 
$-S_{\mu\nu}$ with $S_{\mu\nu}$ given by 
\cite{BI}
\begin{equation} \label{15}
16\pi\,S_{\mu\nu}=2\,\sigma _{(\mu}\,n_{\nu )}-\sigma\,n_\mu\,n_\nu 
-\sigma ^{\dagger}g_{\mu\nu}\ ,
\end{equation}
with
\begin{equation} \label{16}
\sigma _\mu =\sigma _{\mu\nu}\,n^\nu\ ,\qquad \sigma ^{\dagger}
=\sigma _\mu\,n^\mu\ ,\qquad \sigma =g^{\mu\nu}\gamma _{\mu\nu}\ .
\end{equation}
In the present case $\sigma _\mu =0$ and thus $\sigma ^{\dagger}=0$ and 
the surface stress--energy tensor takes the form
\begin{equation} \label{17}
-S_{\mu\nu}=\rho\,n_\mu\,n_\nu\ .
\end{equation}
Hence the energy density of the light--like shell is \cite{BI} 
\begin{equation} \label{18}
\rho =\frac{\sigma}{16\pi}=\frac{1}{4\pi\,r_+^2}\,
\left (m\,k^3-m_+\right )\ .
\end{equation}
Thus the null--cone ${\cal N}$ is the history of a light--like shell 
with isotropic surface stress--energy given by (\ref{17}). We 
note that $m\,k^3$ is the ``mass aspect" of the black--hole, 
in the terminology of Bondi et al.\cite{BBM}, on 
the minus side of ${\cal N}$. A calculation of the singular $\delta$--part of the Weyl tensor 
for $M^-\cup M^+$ reveals that it vanishes.{\it Hence there is 
no possibility of the light--like signal with history ${\cal N}$ containing 
an impulsive gravitational wave}. We note that $\rho$ is 
a monotonically increasing function of $\xi _+$. Thus on the 
interval $-1\leq\xi _+\leq +1$, $\rho$ is maximum at $\xi _+=+1$ (in the direction of the 
motion of the black--hole) and $\rho$ is minimum at $\xi _+=-1$. This 
is as one would expect. A burst of null matter predominantly 
in the direction of motion is required to halt the black--hole. 
In this sense the model we 
have constructed here could be thought of as a limiting case of 
a Kinnersley rocket \cite{K} \cite{B}.

By integrating (\ref{18}) over the shell with 
area element $dA=r_+^2\,d\xi _+\,d\phi _+$ and with $-1\leq\xi _+\leq +1, 
0\leq\phi _+<2\pi$ we obtain the total energy $E$ of the shell 
measured by the distant observer who sees the black--hole, moving 
rectilinearly with 3--velocity  $v$ in the direction $\xi _+=+1$, 
suddenly halted. Thus
\begin{equation} \label{19}
E=\frac{1}{4\pi}\,\int_{0}^{2\pi}d\phi _+\,\int_{-1}^{+1}
\left (m\,k^3-m_+\right )\,d\xi _+\ .
\end{equation}
This results in
\begin{equation} \label{20}
E=m\,\gamma -m_+\ .
\end{equation}
So the energy of the light--like shell is the difference in 
the relative masses of the black--hole before and after the 
emission of the light--like shell. When $v=0$ $(\gamma =1)$ 
the energy of the shell is the difference in the rest--masses 
(naturally taking $m_+<m$) and this is a well--known result 
\cite{BI}. If $v\neq 0$ and $m=m_+$ then
\begin{equation} \label{21}
E=m\,(\gamma -1)\ .
\end{equation}
In this case all of the relative kinetic energy of the black--hole 
before being halted is converted into the relativistic shell.

If the fireball model which we have constructed were to have any 
relevance to the fireball model of gamma ray bursts then (assuming 
that the energy of the relativistically expanding shell 
is converted to gamma rays \cite{PIR}) to produce the expected 
energy $E=10^{52}$ergs using a black--hole of $10^9$ solar masses 
we see from (\ref{21}) that $v=1$km/sec approximately, which is 
one per cent of the speed of our galaxy relative to the Local Group.

We thank Professor W. Israel for helpful discussions and 
Professor T. Piran for encouraging comments.

\end{document}